\documentclass[12pt]{article}
\usepackage{epsfig}
\usepackage{amssymb}
\usepackage{subfigure}

\newcommand{\omd}{\Omega_{\rm d}}
\newcommand{\sa}{\sigma_8}
\newcommand{\omdsf}{\bar{\Omega} _{\rm d} ^{\rm sf}}
\newcommand{\omdn}{\Omega_{\rm d}^0}

\newcommand{\aeq}{a_{\rm eq}}

\newcommand{\wda}{w_{\rm d}}
\newcommand{\wdan}{w_{\rm d}^0}

\def\be{\begin{equation}}
\def\ee{\end{equation}}
\def\bea{\begin{eqnarray}}
\def\eea{\end{eqnarray}}

\begin{document}
\thispagestyle{empty}
\begin{center}
{\Large\bf Quintessence -- the Dark Energy in the Universe?}
\\[2.5cm]
{\large C. Wetterich}\\[.5cm]
{\it Institut f\"ur Theoretische Physik der Universit\"at Heidelberg\\
Philosophenweg 16, D-69120 Heidelberg, Germany}\\[.5cm]
{\small c.wetterich@thphys.uni-heidelberg.de}
\\[2.1cm]

\begin{abstract}
Quintessence -- the energy density of a slowly
evolving scalar field -- may constitute a dynamical form 
of the homogeneous dark energy in the universe. We review
the basic idea and indicate observational tests which may
distinguish quintessence from a cosmological constant.\end{abstract}
\end{center}

\newpage

The idea of quintessence originates from an attempt to understand
the smallness of the ``cosmological constant'' or dark energy in terms of the
large age of the universe \cite{CW1}. As a characteristic consequence,
the amount of dark energy may be of the same order of magnitude as
radiation or dark matter during a long period of the cosmological 
history, including the present epoch. Today, the inhomogeneous energy
density in the universe -- dark and baryonic matter -- is about $\rho_{inhom}\approx(10^{-3} {\rm eV})^4$. This number is tiny in
units of the natural scale given by the Planck mass $M_p=1.22\cdot 10^{19}$ GeV. Nevertheless, it can be understood easily as a direct consequence 
of the long duration of the cosmological expansion: a dominant
radiation or matter 
energy density decreases $\rho\sim M^2_pt^{-2}$ and the
present age of the universe is huge, $t_0\approx 1.5\cdot 10^{10}$ yr.
It is a natural idea that the homogeneous part of the energy density 
in the universe -- the dark energy -- also decays with time and therefore
turns out to be small today\footnote{For some related ideas see ref.
\cite{CC}, \cite{SF}.}.

A simple realization of this idea, motivated by the anomaly of the
dilatation symmetry, considers a scalar field $\phi$ with
an exponential potential \cite{CW1}
\begin{equation}\label{1}
{\cal L}=\sqrt g\left\{\frac{1}{2}\partial^\mu\phi\partial_\mu\phi+
V(\phi)\right\}\ ,\quad V(\phi)=M^4\exp(-\alpha\phi/M)
\end{equation}
with $M^2=M^2_p/16\pi$. In the simplest version $\phi$ couples only to
gravity, not to baryons or leptons. Cosmology is then determined by the
coupled field equations for gravity and the scalar ``cosmon'' field
in presence of the energy density $\rho$ of radiation or matter. 
For a homogeneous and flat universe they read ($n=4$ for radiation and $n=3$ for nonrelativistic matter)
\begin{eqnarray}\label{AA}
&&H^2=\frac{1}{6M^2}(\rho+\frac{1}{2}\dot\phi^2+V),\nonumber\\
&&\ddot\phi+3H\dot\phi+\frac{\partial V}{\partial\phi}=0,\nonumber\\
&&\dot\rho+nH\rho=0.\end{eqnarray}
One finds that independently of the precise initial conditions the 
behavior for large $t$ approaches an exact ``cosmological attractor
solution'' (or ``tracker solution'') where the scalar kinetic and 
potential energy density scale proportional to  matter
or radiation \cite{CW1}
\begin{equation}\label{2}
\phi=\frac{2M}{\alpha}\ln(t/\bar t)\ ,\quad \frac{1}{2}\dot\phi^2
=\frac{2M^2}{\alpha^2}t^{-2}\ , \quad V=\frac{2M^2}{\alpha^2}
\frac{(6-n)}{n}t^{-2},\end{equation}
with the usual decrease of the Hubble parameter $H$
\begin{equation}\label{3}
H=\frac{2}{n}t^{-1}\quad,\quad \rho\sim t^{-2}.\end{equation}
This simple model predicts a fraction of dark energy (as
compared to the critical energy density $\rho_c=6M^2H^2$) which is
constant in time
\be\label{4}
\Omega_d=(V+\frac{1}{2}\dot\phi^2)/\rho_c=
\rho_\phi/\rho_c=\frac{n}{2\alpha^2}\ee
both for the radiation-dominated $(n=4)$ and matter-dominated
$(n=3)$ universe $((\Omega_d+\rho/\rho_c)=1)$.
This would lead to a natural explanation why 
today's dark energy is of the same order of magnitude as dark matter.

The qualitative ingredients for the existence of the stable
attractor solution\footnote{For more details see refs.
\cite{rp}, \cite{clw}, \cite{wet}.} (\ref{2}), (\ref{3}) are
easily understood: for a large value of $ V(\phi)$ the force term
in eq. (\ref{AA}), $\partial V/\partial\phi=-(\alpha/M)V$, is large,
and the dark energy decreases faster than matter or radiation. In
the opposite, when the matter or radiation energy density is much 
larger than $V$, the force is small as compared to the damping term
$3H\dot\phi$ and the scalar ``sits and waits'' until the 
radiation or matter density is small enough such that the overdamped 
regime ends. Stability between the two extreme situations is reached
for $V\sim \rho$.

From present observations one concludes that today's fraction of dark energy
is rather large
\be\label{5}
\Omega^0_d=0.6-0.7.\ee
On the other hand, structure formation would be hindered by a too
large amount of dark energy \cite{fj}, and one infers an approximate
upper bound for the amount of dark energy during structure formation
(for details see below)
\be\label{6}
\Omega^{sf}_d\stackrel{\scriptstyle<}{\sim}0.2.\ee
As a consequence, the fraction of dark energy must have
increased in the recent epoch since the formation of structure.
This implies a negative equation of state for quintessence \cite{cds},
\cite{swz} $w_d=p_\varphi/\rho_\varphi<0$ and can lead to a universe
whose expansion is presently accelerating, as suggested by the 
redshifts of distant supernovae \cite{perl}.

The pure exponential potential in eq. (\ref{1}) is too simple
to account for the recent increase in $\Omega_d$. Possible modifications
of the basic idea of quintessence include the use of other potentials
\cite{CW1}, \cite{rp}, \cite{as}-\cite{bcn}, the coupling of quintessence to
dark matter \cite{wet}, \cite{ame}, nonstandard scalar kinetic terms
\cite{ams} or the role of nonlinear fluctuations \cite{CDM}. We note that
these ideas may not be unrelated, since the presence of large 
fluctuations can modify the effective field equations (e.g. change 
the effective cosmon potential and kinetic term) and lead to a coupling
between quintessence and dark matter \cite{CDM}.

In view of the still very incomplete theoretical understanding of the 
origin of quintessence the choice of an appropriate effective action for the
cosmon is mainly restricted by observation. For comparison with observation
and a discussion of naturalness of various approaches \cite{Heb} we
find it convenient to work with a rescaled cosmon field such that
 the scalar field lagrangian reads
\be
{\cal L}(\varphi)=\frac{1}{2}\,(\partial\varphi)^2\,k^2(\varphi)+
\exp[-\varphi]\,.
\label{A}
\ee
Here and in what follows all quantities are measured in units of the reduced
Planck mass $\overline{M}_P$, i.e., we set $\overline{M}_P^2\equiv M_P^2/(8
\pi)\equiv(8\pi G_N)^{-1}=2M^2=1$. The lagrangian of Eq.~(\ref{A}) contains
a simple exponential potential $V=\exp[-\varphi]$ and a non-standard kinetic
term with $k(\varphi)>0$. If one wishes, the kinetic term can be brought
to the canonical form by a change of variables. Introducing the field
redefinition
\be\label{B}
\phi=K(\varphi) \qquad,\qquad k(\varphi)=\frac{\partial K(\varphi)}
{\partial\varphi}
\ee
one obtains
\be\label{C}
{\cal L}(\phi)=\frac{1}{2}\,(\partial\phi)^2+\exp[-K^{-1}(\phi)]\,.
\ee
The exponential potential in eq. (\ref{1}) corresponds to a constant
\be\label{D}
k=\frac{1}{\sqrt2\alpha}\ee

We restrict our discussions to potentials that are monotonic in $\phi$.
(Otherwise, the value of the potential at the minimum must be of the order
of today's cosmological constant, with $V_{min}\approx 10^{-120}$.
Cosmologies of this type are discussed in~\cite{as}, \cite{saw}.) All monotonic
potentials can be rescaled to the ansatz Eq.~(\ref{A}). An initial value
of $\varphi$ in the vicinity of zero corresponds then to an initial scalar
potential energy density of order one. We consider this as a natural
starting point for cosmology in the Planck era. As a condition for
naturalness we postulate that no extremely small parameter should be
present in the Planck era. This means, in particular, that $k(0)$ should be
of order one. Furthermore, this forbids a tuning to many decimal places of 
parameters appearing in $k(\varphi)$ or the initial conditions. For natural
quintessence all characteristic mass scales are given by $\overline{M}_P$
in the Planck era. The appearance of small mass scales during later
stages of the cosmological evolution is then a pure consequence of the
age of the universe (and the fact that $V(\varphi)$ can be arbitrarily close to
zero). In addition, we find cosmologies where the late time behaviour is
independent of the detailed initial conditions particularly attractive. For
such tracker solutions \cite{CW1,wet,rp,clw,cds} no detailed understanding of the dynamics in the Planck era is needed. 
It is indeed possible to find \cite{Heb} viable cosmological solutions with high present-day
acceleration which are based on functions $k(\varphi)$ that always remain
${\cal O}(1)$.

It is convenient to analyse the cosmological evolution using the scale
factor $a$ instead of time as the independent variable. In this case, the
evolution of matter and radiation energy density is known explicitly and
one only has to solve the set of the two differential equations for the
homogeneous dark energy density $\rho_\varphi$ and the cosmon field $\varphi$
\be\label{E}
\frac{d \ln \rho_\varphi}{d \ln a}= -3(1+w_\varphi)\,\,\,,\qquad
\frac{d \varphi}{d \ln a}=\sqrt{6 \Omega_T/k^2(\varphi)}\,\,,
\ee
with $\Omega_T=T/(3H^2)$ the fraction of kinetic field energy and $w_d=
p_\varphi/\rho_\varphi$. Here the cosmon kinetic energy is denoted by $T=
\dot{\varphi}^2k^2(\varphi)/2$ whereas $p_\varphi=T-V$ and $\rho_\varphi=T+V$ specify
the equation-of-state of quintessence. Thus, more explicitly, the cosmology
is governed by four equations for the different components of
the energy density $\rho_m,\rho_r,\rho_\varphi$ and $\varphi$
\bea
\frac{d\ln\rho_m}{d \ln a}=-3\,(1+w_m)\,\,\,,\hspace*{0.8cm}
&&\frac{d\ln\rho_r}{d \ln a}=-3\,(1+w_r)\,\,,
\nonumber\\ \label{F}\\
\frac{d \ln \rho_\varphi}{d \ln a}= -6\left(1-\frac{V(\varphi)}{\rho_\varphi}
\right)\,\,\,,&&\frac{d\varphi}{d \ln a}=\sqrt{\frac{6\,(\rho_\varphi-
V(\varphi))}
{k^2\,(\varphi)(\rho_m+\rho_r+\rho_\varphi)}}\,\,,\nonumber
\eea
where $w_m=0$ and $w_r=1/3$ for matter and radiation respectively.
For our exponential potential $V=\exp[-\varphi]$, the last equation can be 
rewritten as 
\be
\frac{d\ln V}{d \ln a}=-\sqrt{\frac{6\,(\rho_\varphi-V)}
{k^2\,(-\ln V)(\rho_m+\rho_r+\rho_\varphi)}}\,.\label{G}
\ee
We note that today's value of $\rho_\varphi$ plays the role of $\epsilon
_{vac}$ and the fraction of dark energy is therefore
$\Omega_d=\rho_\varphi/(3H^2)$. For a rough
orientation, today's value of $\varphi$ must be $\varphi_0\simeq 276$ for
all solutions where the present potential energy is of the
order of $\epsilon_{vac}$.

The simplest case, $k(\varphi)=k=$ const., (cf. eq. (\ref{D})) 
corresponds to the original
quintessence model~\cite{CW1}. If
$k^2<1/n$ (with $n=3(1+w_{r,m})$
for radiation and matter domination, respectively), then
the scalar field energy $\rho_\varphi$ follows the evolution of the background
component $\rho$ in the way described above, with
$\Omega_d=nk^2$.
This attractor solution can be easily retrieved from Eqs.~(\ref{F}) and 
(\ref{G}) by noting 
the constancy of $\rho_\varphi/\rho$ and $V/\rho$. For $k^2>1/n$ the 
cosmological attractor is a scalar dominated universe \cite{
CW1,wet,clw,pli} 
with $H=2 k^2t^{-1},\ w_d=1/(3k^2)-1$. However, it
has been emphasized early~\cite{CW1} that there is actually no reason
why $k(\varphi)$ should be exactly constant and that interesting cosmologies
may arise from variable $k(\varphi)$. In particular, one may imagine an
effective transition from small $k$ (small $\Omega_d$) in the early
universe (nucleosynthesis etc.) to large $k$ ($\Omega_d\simeq 1$)
today~\cite{wet,swz,bcn,mp}.

\begin{figure}[ht]
\begin{center}
\vspace*{.2cm}
\parbox[b]{15.7cm}{\psfig{width=15.7cm,file=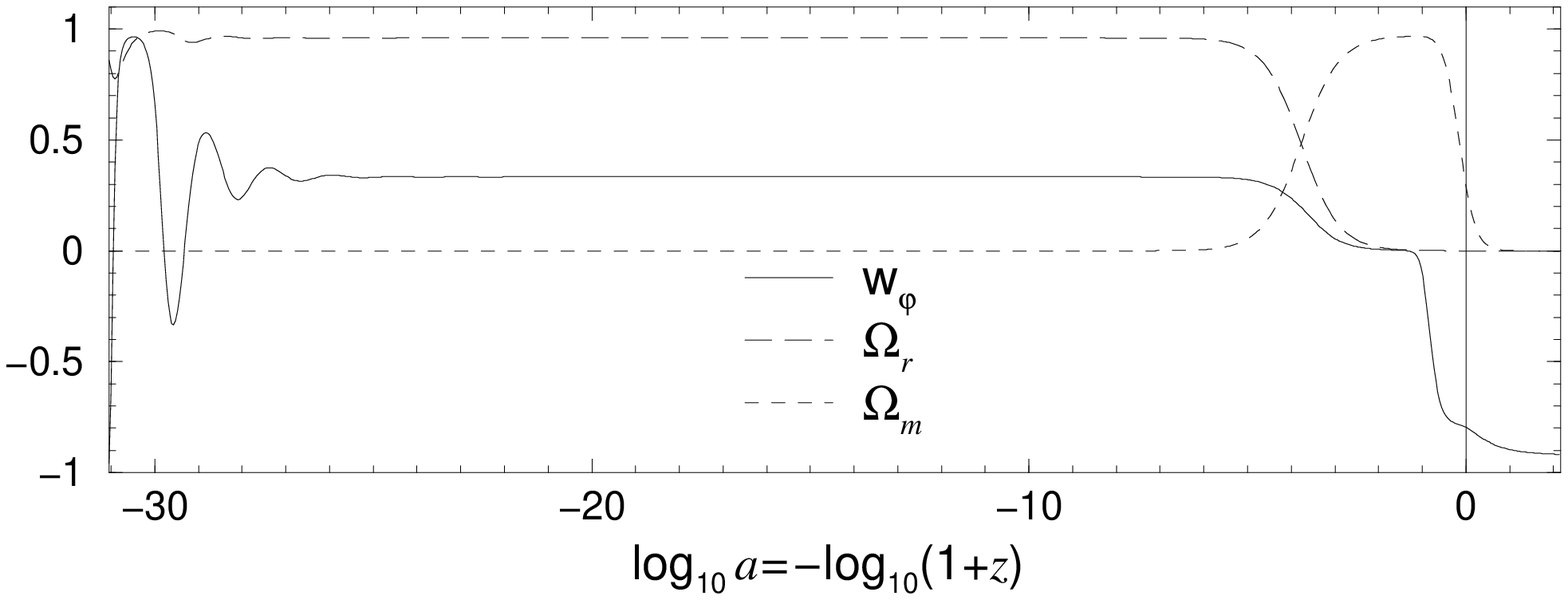}}\\
\end{center}
\refstepcounter{figure}\label{jump}
{\bf Figure \ref{jump}:} Cosmological evolution with a leaping kinetic term.
We show the fraction of energy in radiation ($\Omega_r$) and matter
($\Omega_m$) with $\Omega_d=1-\Omega_r-\Omega_m$. The equation of state
of quintessence is specified by $w_\varphi$.
\end{figure}

A particularly simple case of a $\varphi$ dependent kinetic coefficient
$k(\varphi)$ is obtained if $k$ suddenly changes from a small number $k<0.22$
(consistent with nucleosynthesis and structure formation bounds) to a
number above the critical value $1/\sqrt{n}$. Consider, for example, the
function
\be
k(\varphi)=k_{min}+\mbox{tanh}(\varphi-\varphi_1)+1\qquad\qquad
(\mbox{with}\quad  
k_{min}=0.1\,,\,\,
\,\varphi_1=276.6\,)\,,\label{H}
\ee
that gives rise to the cosmological evolution of Fig.~\ref{jump}. This
``leaping kinetic term quintessence''
model, which completely avoids the explicit use of very large or very small
parameters, realizes all the desired features of quintessence \cite{Heb}. The 
homogeneous dark energy density tracks below the background component in 
the early universe
($k=0.1$) and then suddenly comes to dominate the evolution when $k$ rises
to a value $k=2.1$ approximately today. With a tuning on the percent level
(the value of $\varphi_1$ has to be appropriately adjusted) realistic
present-day values of $\Omega_d$ and $w_d^0$ can be realized. In the
above example, one finds $\Omega_d^0=0.70$ and $w_d^0=-0.80$.
Note that, due to the extended tracking period, the late cosmology is
completely insensitive to the initial conditions. In the example of
Fig.~\ref{jump}, the evolution starts at the Planck epoch with a total
energy density $\rho_{tot}=1.0$, $\varphi=2.0$ and $\dot{\varphi}=0$
(corresponding to $\Omega_d=0.14)$. We have checked explicitly other
initial conditions, e.g., with $\Omega_d$ near one. 
The present day value $w_d^0$ can be forced to be even closer to $-1$ if
the leap of $k(\varphi)$ is made sharper or the final value of $k$ is made
higher by a simple generalization of Eq.~(\ref{H}). Thus, all scenarios
between a smoothly rising quintessence contribution and a suddenly emerging
``cosmological constant'' can be realized.

As the theoretical understanding of the origin of quintessence from 
fundamental physics remains very incomplete, a large variety of effective
potentials and kinetic terms for the cosmon can be conceived. One would
therefore like to use the available information from observation
to determine the characteristic features of quintessence in a way
that is as model-independent as possible. The basic feature which
distinguishes quintessence from a cosmological constant is the time
evolution of the dark energy. For a cosmological constant the energy
density is constant, and therefore $\Omega_d\sim t^2$ becomes
irrelevant in the early universe. In contrast, the time evolution
of $\Omega_d(t)$ is more complex for quintessence. In particular,
a relevant fraction of the energy density may have been dark energy also
in earlier epochs of the universe. The effects of this ``early dark
energy'' may lead to observable consequences. We therefore aim to
gather information about the value of $\Omega_d(t)$ at
various characteristic moments of the cosmological evolution. As an alternative to an overall fit of the data, which typically involves
many cosmological parameters and has to be done, in principle,
for a large variety of different quintessence models, we pursue here
a search for ``robust quantities'' that can ``measure'' $\Omega_d(t)$ for
different $t$. Typical examples are the determination of the present 
fraction in dark energy $\Omega_d^0\approx 0.6-0.7$
or the  bound from nucleosynthesis \cite{CW1}, \cite{bs} $\Omega_d^{ns}
\stackrel{\scriptstyle<}{\sim} 0.2$.

As an example we discuss here how the amount of dark energy at the 
time of last scattering, $\Omega_d^{ls}$, may be extracted from
cosmic microwave background (CMB) anisotropies. Recent measurements
of the CMB \cite{B1, B2}
show three peaks as distinct features, seeming to confirm beyond any
reasonable doubt the inflationary picture of structure formation from
predominantly adiabatic initial conditions. 
It was demonstrated \cite{DL1, DL2, DL3} that the location of the CMB
peaks depends on three dark-energy related quantities: the amounts of
dark energy today $\Omega^0_{d}$ and at last scattering
$\overline{\Omega}_{d}^{\rm _{\, ls}}$ as well as its time-averaged
equation of state $\overline{w}_0$.   

The CMB peaks arise from acoustic oscillations of the primeval plasma
just before the universe becomes translucent.  The angular momentum
scale of the oscillations is set by the \emph{acoustic scale} $l_A$
which for a flat universe is given by 
\begin{equation}
\label{I}
l_A = \pi \frac{\tau_0 - \tau_{\rm ls}}{\bar c_s \tau_{\rm ls}},
\end{equation}
where $\tau_0$ and $\tau_{\rm ls}$ are the conformal time today and at
last scattering and $\bar{c}_s$ is the average sound speed before
decoupling.  The value of $l_A$ can be calculated simply, and for flat
universes is given by \cite{DL1}
\begin{equation}
\label{J}
 l_A = \pi \bar c_s^{-1} \Bigg[
      \frac{F(\Omega^0_{d},\overline{w}_0)}{(1-{\overline{\Omega}^{\rm
      _{\, ls}}_{d}})^{1/2}} \Bigg \{ \left({a_{\rm ls} +
      \frac{\Omega^0_{\rm r}}{ 1 - \Omega^0_{d}}}\right)^{1/2}  -
      \left({\frac{\Omega^0_{\rm r}}{1 -
      \Omega^0_{d}}}\right)^{1/2} \Bigg \} ^{-1} - 1 \Bigg],
\end{equation}  
 The conformal time
\be\label{J1}
\tau_0=2H_0^{-1}(1-\Omega^0_d)^{-1/2}F(\Omega^0_d,\bar w_0)\ee
involves the integral
\begin{equation} \label{K}
 F(\Omega_{d}^0,\overline{w}_0) = \frac{1}{2} \int_0^1
\textrm{d}a \Bigg( a + \frac{\Omega_{d}^0}{1-\Omega_{d}^0} \, a ^{(1
- 3 \overline{w}_0)}  + \frac{\Omega_{\rm r}^0(1-a)}{1-\Omega_{d}^0}
\Bigg)^{-1/2}.
\end{equation} 
Here $\Omega^{0}_{\rm r}, \Omega^{0}_{d}$ are today's radiation
and quintessence components, $a_{\rm ls}$ is the scale factor at last
scattering (if $a_0=1$), $\bar c_s,\overline{\Omega}^{\rm _{\,
ls}}_{d} $ are the average sound speed and quintessence components
before last scattering \cite{DL1} and $\overline{w}_0$ is the
$\Omega_d$-weighted equation of state of the Universe
\begin{equation}
\label{L}
\overline{w}_0 = \int_0^{\tau_0} \Omega_{d}(\tau) w_d(\tau) \textrm{d}
\tau \times \left( \int_0^{\tau_0} \Omega_{d}(\tau) \textrm{d} \tau
\right)^{-1}.
\end{equation}

The location of the peaks is influenced by driving effects and
we compensate for this by parameterising the location of the $m$-th
peak $l_m$ as in \cite{Hu}
\begin{equation} \label{M}
l_m \equiv l_A \left(m - \varphi_m\right). 
\end{equation}
The reason for this parameterization is that the phase shifts
$\varphi_m$ of the peaks are determined predominantly by
pre-recombination physics, and are independent of the geometry of the
Universe. The values of the phase shifts are typically in the range
$0.1 \dots 0.5$ and depend on the cosmological parameters $\Omega_b
h^2, n, \overline{\Omega}^{\rm _{\, ls} }_{\varphi}$ and the ratio of
radiation to matter at last scattering $r_\star = \rho_r(z_\star) /
\rho_m(z_\star).$ 

It was shown \cite{DL2} that $\varphi_3$ is relatively
insensitive to cosmological parameters, and that by assuming the
constant value $\varphi_3 = 0.341$ we can estimate $l_A$ to within one
percent if the location of the third peak $l_3$ is measured, via the
relation $l_A=l_3/(3-\varphi_3)$.
The measurement of a third peak in the CMB spectrum by BOOMERANG
\cite{B1} now allows us to extract the acoustic scale
$l_A$ and use this as a constraint on cosmological models. 
(See Fig. 2). From the conservative assumption that
$800<l_3<900$,one gets the bound
\be\label{N}
296\leq l_A\leq 342\ee
(The BOOMERANG analysis \cite{B3} indicates $l_A=316\pm8$.)
For a given value of the Hubble parameter $h=0.65$ and present dark
energy $\Omega_d^0=0.6$ one finds \cite{DL1} for the ``leaping kinetic
term quintessence'' discussed above  with $\Omega_d^{ls}=0.15$ a value
$l_A=300$ whereas a cosmological constant yields $l_A=296$.
On the other hand, power law quintessence \cite{rp} leads to $l_A=270$. 
Larger values of $\Omega^0_d$ increase $l_A$ (cf. eq. (\ref{J})) and
similar for larger values of $h$. Using constraints on $\Omega^0_d$ and
$h$ from other observations a distinction between different quintessence
models becomes possible already with the present data \cite{DL3}.
For example, power law potentials seem to be disfavored. 

\bigskip
\bigskip

 \begin{figure}[h]
 \begin{center}
   \epsfig{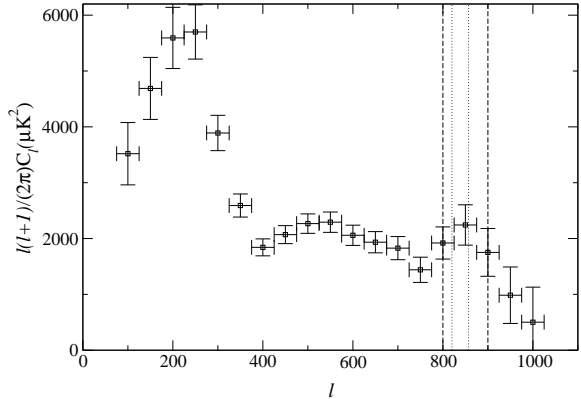} \caption{The CMB
   anisotropy power spectrum as measured by BOOMERANG [23].   
   The inner vertical lines show the
   region $820<l_3<857$ as calculated by the BOOMERANG team [29],
   and the outer lines our more conservative region
   $800<l_3<900$.\label{fig:boom}}
 \end{center}
 \end{figure}

Another possible benchmark for determining 
the amount of dark energy in early cosmology is the formation
of structure \cite{JS}.
In particular, $\sa$, the rms density 
fluctuation averaged over $8 h^{-1} \textrm{Mpc}$ spheres, is a
sensitive parameter. 
The COBE \cite{CO} normalization \cite{CNN} 
of the CMB power spectrum determines $\sa$ for 
any given model by essentially fixing the fluctuations at decoupling. 
This prediction is to be compared to values of $\sa$ infered from other observations, such as 
cluster abundance constraints which yield \cite{WS}
\begin{equation}\label{P}
 \sigma _8 =(0.5 \pm 0.1)\Omega _m^{-\gamma},
\end{equation} 
where $\gamma$ is slightly model dependent and usually  $\gamma\approx 0.5$.
A model where these two $\sa$ values do not agree can be ruled out.
Standard Cold Dark Matter (SCDM)  without dark energy
\footnote{and $h=0.65,\ \hat n=1, \Omega_{\rm b}h^2 =0.021,\ \Omega_{\rm m}^0=1$}
 for instance gives  $\sa^{\rm cmb} \approx 1.5,\ \sa^{\rm clus.} \approx 0.5 \pm 0.1$  
and is hence incapable of meeting both constraints.

Within the standard scenario\footnote{See \cite{CDM} for an alternative
with cosmon dark matter.} where structure formation proceeds by
the gravitational clumping of cold dark matter one can estimate
\cite{JS}
the CMB-normalized $\sa$-value for 
a very general class of 
quintessence models Q just from the knowledge of  the ``background solution'' 
$[\omd(a),\ \wda(a)]$ 
and the $\sa$-value of the $\Lambda$CDM model $\Lambda$ with 
the same amount of dark energy today given by a cosmological
constant, i.e. with 
 $\Omega _{\Lambda}^0=\omdn$:
\begin{equation} \label{Q}
 \frac{\sigma _8 (Q)}{\sigma _8 (\Lambda)}\approx
\left( a _{\rm eq}\right)^{ 3\, \omdsf / 5}
 \left(1-\Omega _{\Lambda}^0 \right)^{-\left (1+ \bar w ^{-1}\right)/5}
 \sqrt{\frac{\tau _0 (Q)}{\tau _0 (\Lambda)}}.
\end{equation} 
If Q is a model with `early quintessence', $\omdsf$ is an average value 
for the fraction of dark energy 
during the matter dominated era, before $\omd$ starts growing rapidly at 
scale \mbox{factor $a_{\rm tr}$:}
\begin{equation}\label{R}
 \omdsf \equiv [ \ln{a_{\rm tr}}-\ln{\aeq} ]^{-1} \int_{\ln \aeq}^{\ln a_{\rm tr}} \omd(a)\  {\rm d} \ln a  .
\end{equation}
(For a model without early quintessence, $\omdsf$ is zero.) The effective equation of state of quintessence
 $\bar{w}$,  is an
average value for $\wda$ during the time in which $\omd$ is growing rapidly:
\begin{equation}\label{S}
\frac{1}{\bar{w}}=\frac{\int _{\ln a_{\rm tr}}^0 \omd(a)/w(a)\:  d \ln{a}}
                   {\int _{\ln a_{\rm tr}}^0 \omd(a)\:  d  \ln{a}}.
\end{equation}
In many cases, the present equation of state, 
$\wdan$,  can be used as an approximation to $\bar{w}$ since the
integrals are dominated by periods with large $\omd$. (In general,
$\bar w\not=\bar w_0$, Eq. (\ref{L}).)
The scale factor at
matter radiation equality is
\begin{equation}\label{T}
  \aeq=\frac{\Omega _{\rm r}^0}{\Omega _{\rm m}^0}= \frac{4.31 \times 10^{-5}}
  {h^2(1-\omdn)}.
\end{equation}
Finally, 
the conformal age of the universe $\tau_0$ is given by eq. (\ref{J1}).
\mbox{Equation (\ref{Q})}  in combination with
(\ref{P}) can be used to make general statements about the consistency of quintessence
models with $\sa$-constraints.

The CMB-normalized value of $\sa$ 
depends on all
cosmological parameters. As a rough guide for the strength of these dependencies 
around standard values $\omdn = 0.65$, $h=0.65$, spectral
index $\hat n=1$,
$\Omega _b h^2=0.02$ with  $-1<\bar w<-0.5$ we get
\begin{itemize}
\item Increasing $h$ by 0.1 $\Rightarrow$ Increase of $\sigma _8$ by 20 \%
\item Increasing $\omdn$ by 0.1 $\Rightarrow$
      Decrease of $\sigma _8$ by 20\%   
\item Increasing $\hat n$ by 0.1 $\Rightarrow$ Increase of 
      $\sigma _8$ by 25\%
\item Increasing $\bar w$ by 0.1 $\Rightarrow$ Decrease of $\sa$ by 5-10\%
\item Increasing $\Omega _b h^2$ by 0.01 $\Rightarrow$
      Decrease of $\sigma _8$ by 10\%
\item Increasing $\omdsf$ by 0.1 $\Rightarrow$
      Decrease of $\sigma _8$ by 50\%
\end{itemize}
Comparing with observation, the dependencies listed can be used for a
quick check of viability for a given quintessence model and parameter
set.  
If $\omdn$ is increased by $0.1$, cluster abundances according to
Eq. (\ref{P})  yield an approx. $20 \%$ higher value of $\sa^{\rm cluster}$.
In combination with the corresponding decrease of $\sa^{\rm cmb}$,
the net effect on the ratio $\sa^{\rm cmb} / \sa^{\rm cluster}$  
is therefore a decrease by $33 \%$.
For a $\Lambda$CDM universe with standard values as above one has $\sa^{\rm cmb} =
0.90$ and $\sa^{\rm cmb} / \sa^{\rm cluster} = 1.01 \pm 0.2$.
Compatibility of the cosmological scenario requires this ratio to be
close to unity.

In particular, we note the degeneracy between the amount of dark
energy during structure formation $\bar\Omega^{sf}_d$ and the spectral
index $\hat n$, as shown in fig. 3. In this context it may be useful
to recall that the smallness of the density fluctuations could find
a natural explanation within inflation without invoking small parameters
or mass ratios if $\hat n\approx 1.15$ \cite{CWDF}.

Constraints from structure formation may be combined with the CMB-data to
constrain specific models of quintessence \cite{DL3}. For a specific
quintessence model one may further use the supernovae results \cite{perl}.
They constrain the very recent increase of $\Omega_d$, or, 
equivalently, the present equation of state $w^0_d$. For leaping kinetic
term quintessence the value of $w^0_d$ is, however, not directly related 
to $\Omega^{ls}_d$ or $\Omega^{sf}_d$ since it is strongly
influenced by the width and height of the leap (generalizing eq. (\ref{H})
). We conclude that some particular models of quintessence are disfavored
already by the present data, as, for example, power law potentials
\cite{DL3}. Other models, like leaping kinetic term quintessence
\cite{Heb}, are consistent with present observations, allowing early
quintessence on a level of $\Omega_d\stackrel{\scriptstyle<}{\sim}0.2$
during nucleosynthesis, structure formation and last scattering. With
forthcoming observations it should be possible to improve the 
constraints substantially and to distinguish between 
quintessence and a cosmological constant.

Observation seems to tell us that for a viable model of quintessence the
fraction of dark energy $\Omega_d$ has increased substantially in 
recent time, say from 0.1 to 0.7. Even though an increase by a factor 
of 7 since the Planck epoch is mild as compared to the factor $\sim 10^{120}$
for a cosmological constant, it has happened during a relatively short 
cosmological period and singles out the present epoch. So far, 
phenomenological models of quintessence can deal with this by
adjusting a suitable parameter on the percent level. We feel, however,
that this answer to the question ``why now'' is 
not very satisfactory. In our view, quintessence would become more 
credible if the recent increase  in $\Omega_d$ can be induced by
a recent characteristic event in cosmology, like the formation 
of structure. Cosmon dark matter \cite{CDM} could give such an
explanation. It would also lead to a revision of the present
picture of cold dark matter and structure formation. We should 
not be surprised if quintessence leads us to further changes of our 
picture of the universe!

\begin{figure*}[!ht]
\begin{center}
\subfigure[$h=0.7,\ \omd^0=0.6$]{ \label{forCmp}
\includegraphics[scale=0.38, angle =-90]{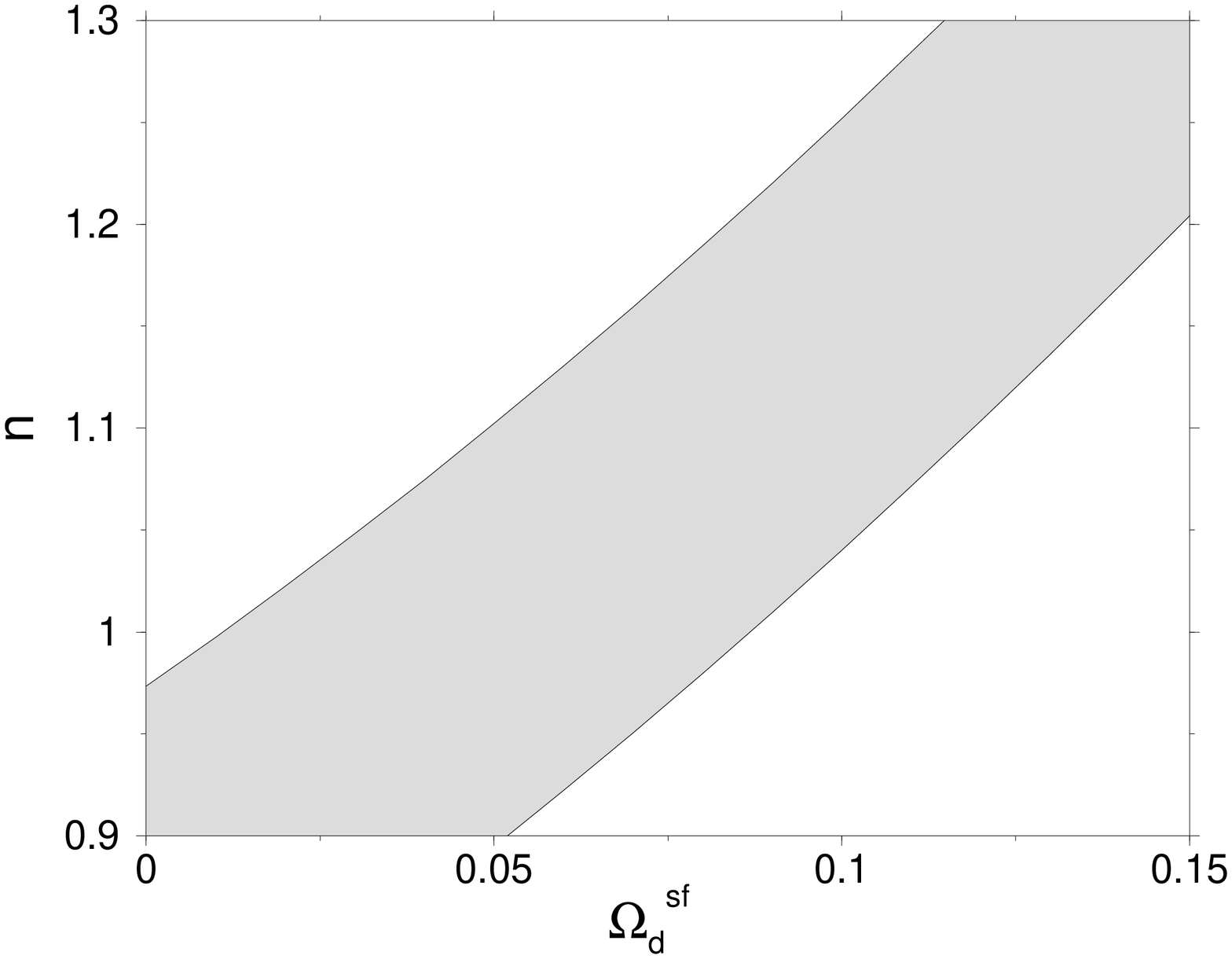} }
\end{center}
\caption{Allowed range of early quintessence and spectral index
$\hat n$ for given  values of the Hubble parameter $h$ and present
dark energy $\omd^0$.}
\label{range}
\end{figure*}

\noindent{\bf Acknowledgement:} The author would like to thank
M. Doran, A. Hebecker, M. Lilley, and M. Schwindt for collaboration
on the content of this work. Part of it is based on refs. \cite{Heb},
\cite{DL3}, \cite{JS}.

\end{document}